\documentclass[conference,10pt]{IEEEtran}
\usepackage{epsfig,epsf,rotating,setspace,latexsym,amsmath,amssymb,amsfonts,bm,theorem,subfigure,epstopdf}
\usepackage{cite,authblk}
\usepackage{bbm}
\usepackage{algorithm}
\usepackage[noend]{algpseudocode}
\usepackage{color}
\usepackage{mathtools}
\usepackage{soul}
\DeclareMathOperator*{\argmax}{arg\,max}

\algrenewcommand\algorithmicforall{\textbf{foreach}}
\algrenewcommand\algorithmicindent{.8em}

\definecolor{green1}{rgb}{0.2,0.7,0.2}

\newtheorem{lemma}{Lemma}

\newenvironment{Proof}[1]{\medskip\par\noindent{\bf Proof:\,}\,#1}{{\mbox{\,$\blacksquare$}\par}}

\IEEEoverridecommandlockouts
\allowdisplaybreaks

\begin{document}

\title{Optimizing Profitability in Timely Gossip Networks \thanks{Research of UIUC authors was supported in part by the ARO MURI Grant AG285 and the AFOSR Grant FA9550-19-1-0353.}}

\author[1]{Priyanka Kaswan}
\author[2]{Melih Bastopcu}
\author[1]{Sennur Ulukus}
\author[2]{S. Rasoul Etesami}
\author[2]{Tamer Ba{\c s}ar}

\affil[1]{\normalsize Department of Electrical and Computer Engineering, University of Maryland, College Park, MD 20742}
\affil[2]{\normalsize Coordinated Science Laboratory, University of Illinois Urbana Champaign, Urbana, IL-61801}

\maketitle

\begin{abstract}
We consider a communication system where a group of users, interconnected in a bidirectional gossip network, wishes to follow a time-varying source, e.g., updates on an event, in real-time. The users wish to maintain their expected version ages below a threshold, and can either rely on gossip from their neighbors or directly subscribe to a server publishing about the event, if the former option does not meet the timeliness requirements. The server wishes to maximize its profit by increasing subscriptions from users and minimizing event sampling frequency to reduce costs. This leads to a Stackelberg game between the server and the users where the sender is the leader deciding its sampling frequency and the users are the followers deciding their subscription strategies. We investigate equilibrium strategies for low-connectivity and high-connectivity topologies. 
\end{abstract}

\section{Introduction} \label{sec:intro}
We consider a setting with a time-varying process or event ($E$) over a discrete timeline that gets updated with a new version of information in every timeslot with probability $p_e$, i.e., the inter-update times of the event are distributed as geometric with $p_e$. Denoting the current version of the event in timeslot $t$ by $V_E(t)$, each event update increments $V_E(t)$ by one. A set of user nodes $\mathcal{N}$, with cardinality $|\mathcal{N}|$, aim to track the event as timely as possible. Users are connected in a gossip network represented through an undirected graph $ G =(\mathcal{N},E)$ with edge set $E$, such that user $i$ transmits an update packet to user $j$ with probability $p$ in each timeslot if there is an edge between them, i.e., $e_{ij}\in E$; see Fig.~\ref{fig:twoway_line_network} and Fig.~\ref{fig:fully_connected_subscriber} for examples. We quantify timeliness at the users through the version age metric \cite{aoi_gossip_litsurvy, Yates21gossip, kaswan_jamming_jrl, baturalp21comm_struc, mitra_srcrate_learning24}, which is a natural choice in this setting given the versioning of information. Denoting the current version at an arbitrary user node $i$ in timeslot $t$ by $V_i(t)$, the instantaneous version age at user $i$ is then $X_i(t)=V_E(t)-V_i(t)$. A service provider, e.g., a server, samples the event with probability $\beta$ in each timeslot $t$ and publishes the latest sampled version of information. 

To stay ``reasonably updated'' about the event, a notion that we formalize later, a user can explore two avenues: it can either rely on the neighbors from the underlying gossip network for updates, or it can directly subscribe to the server if the former option cannot meet the desired timeliness level at the user. This setting holds significant relevance in the upcoming hyper-connected 5G/6G networks where IoT service providers and newswires constantly monitor dynamic data sources to offer the latest updates to users in exchange for a nominal fee \cite{zhang_pricingfreshdata}. 

This model was studied in \cite{kaswan_subscription_direct} for directed networks such as line, tree and star networks, where updates took fixed directed paths from the source to the users. In \cite{kaswan_subscription_direct}, each user node accepts all packets sent by its predecessor node, since in directed paths, nodes that are fewer hops away from the source have lower instantaneous version ages at all times than users larger hops away. This paper serves as an extension to \cite{kaswan_subscription_direct} by probing into a more generalized case, where information flow is bidirectional between every pair of neighbors. In bidirectional networks, user nodes simultaneously receive packets from multiple nodes, all of which need not have a fresher version than the receiving user, causing the user to discard some of the arriving packets to keep a low version age.

In this respect, we investigate how users will choose to subscribe to the server in bidirectional line networks, representing the most constrained extreme of connectivity spectrum, and in fully-connected networks, representing the most deeply connected extreme of connectivity spectrum. Further, the server in this work is interested in sustaining a desired fraction of subscribing users from the population while minimizing sampling costs to maximize its overall profit. The inherent asymmetry in objectives between the server and the users gives rise to a non-cooperative game between them, formulated as a Stackelberg game. We next formally describe the game formulation, and investigate Stackelberg equilibrium strategies for both the low-connectivity and the high-connectivity topologies. 

\section{Age Analysis and Game Formulation}
All event updates and inter-node transmissions are assumed to consume the entire duration of timeslot and are available to the receiving/processing node by the beginning of the next timeslot. Therefore, if the server samples the source in timeslot $t$, its version number resets to $V_R(t+1)= V_E(t)$ at the beginning of the following timeslot. Let $U_{i,j}(t)$ indicate whether an update packet is transmitted from node $i$ to node $j$ at time $t$: $U_{i,j}(t)=1$ implies that an update was sent 
and $U_{i,j}(t)=0$ implies that no packet was sent. We define $U_{E}(t)$ in an analogous fashion to indicate occurrence of an event update. When $U_{E}(t)=1$, since the event update consumes an entire timeslot, it cannot be forwarded to the server in timeslot $t$ itself upon sampling, causing instantaneous version age at server $X_R(t+1)$ and users to increment by one.

\subsection{Average Age Analysis at the Server}
Instantaneous version age at the server evolves as \cite[(1)]{kaswan_subscription_direct}
\begin{align}\label{eqn:instantage_server}
    X_R(t+1) = & 0 \times (1-U_{E}(t))U_{E,R}(t) 
    + 1 \times U_{E}(t)U_{E,R}(t) \nonumber\\
    & + X_R(t) \times (1-U_{E}(t))(1-U_{E,R}(t))\nonumber\\
    & + (X_R(t)+1)\times U_{E}(t)(1-U_{E,R}(t)),
\end{align}
where $X_R(t+1)$ depends on prior age $X_R(t)$, and transitions $U_{E,R}(t)$ and $U_{E}(t)$.\footnote{(\ref{eqn:instantage_server}) together with some initial distributions can be shown to define a discrete-time Markov chain with state space $\mathbb{Z}_{+}$ that is irreducible, aperiodic and positive recurrent. We omit this analysis here since it is non-instructive.} Defining $x_i=\lim_{t \to \infty} \mathbb{E}[X_i(t)]$  and taking the long-term expectation on both sides of (\ref{eqn:instantage_server}) gives
\begin{align}\label{eqn:longtermexp_age_server}
    x_R =&  p_e\beta + x_R(1-p_E)(1-\beta) + (x_R+1)p_e(1-\beta),
\end{align}
which simplifies to $x_R \!=\!  p_e \beta^{-1}$. Likewise, age at a subscribing user, denoted by $S$, at time $t \! + \! 1$ depends on prior age $X_S(t)$ and transitions $U_E(t)$ and $U_{R,S}(t)$, and as shown in \cite[(4)]{kaswan_subscription_direct},
\begin{align}\label{eqn:longtermexp_age_subscriber}
    x_S= x_R+p_e = p_e\left(\beta^{-1}+1\right).
\end{align}
Note that $U_{R,S}(t)=1$ for all $t$, since the latest version at the server is transmitted to all subscribers in every timeslot. 

\textit{Subscription condition}: User $i$ prefers a bounded age $x_i$, 
\begin{align}\label{eqn:ACconstraint}
    x_i<Lx_S,\quad L> 1,
\end{align}
which we will call the \emph{age compatibility (AC) constraint}. If (\ref{eqn:ACconstraint}) holds without subscription, user $i$ will not subscribe to the server, i.e., its subscription action will be $a_i =0$; otherwise, user $i$ will subscribe to the server, i.e., $a_i =1$. Then, we say that a user is AC-stable if the user ($i$) is a non-subscriber and its expected age satisfies (\ref{eqn:ACconstraint}); or ($ii$) is a subscriber and the alternate decision to unsubscribe while keeping other users' decisions unchanged would cause its expected age to violate (\ref{eqn:ACconstraint}) \cite{kaswan_subscription_direct}. Thus, \textit{AC-stable} users do not change their decisions.

\subsection{Game Formulation between the Server and Users}
The server's goal is to maximize the fraction of subscribers in the population, $F_S=\big(\sum_{i=1}^{|\mathcal{N}|}a_i\big)/|\mathcal{N}|$, to boost its earnings while minimizing the cost of sampling $c(\beta)$, an increasing function of $\beta$. Thus, the utility function of the server given $\beta$ and the action vector $\bm{a}=[a_1,a_2,\cdots, a_{|\mathcal{N}|}]^T\!$, is expressed as $L_R(\beta, \bm{a}) = F_S- c(\beta)$. Users base their actions on their ability to satisfy the AC constraint (\ref{eqn:ACconstraint}). The server's information structure is $I_R \!=\!\!\{G, p_e,p\}$, while each user $i\!\in \!\mathcal{N}$ has $I_i\!=\!\{G,$ $p_e,p, \beta, \bm{a}_{-i}\!\}$ where $\!\bm{a}_{-i}\!$ is actions of the users other than $i$.

The server commits to a strategy $\eta_R:I_R\rightarrow [0,1]$ and each user $i$ adopts a strategy $\gamma_i:I_i\rightarrow\{0,1\}$. Specifically, in this context, user $i$'s action $a_i\!=\!\gamma_i(I_i) \!=\! \mathbf{1}\{ x_{i,ns}\!\!>\!\! L x_S \}$, where $x_{i,ns}$ represents the average age when the user does not subscribe to the server, given $\beta$ and $\gamma_{-i}(\beta)$. The server's utility function is $J_R(\eta_R;\gamma_1, \gamma_2,\ldots, \gamma_{|\mathcal{N}|} ) = L_R(\beta, \bm{a})$, where actions are determined by $ a_i \!\!=\!\!\gamma_i(I_i) $ and $ \beta \!\!=\!\!\eta_R(I_S) $. We consider a Stackelberg game, where the server commits to a strategy $\eta_R$ first, and then the users take action to satisfy the AC condition in (\ref{eqn:ACconstraint}). We say that the set of strategies $(\eta_R^*,\{\gamma_i^*(\eta_R^*,\gamma_{-i}^*)\}_{i=1}^{|\mathcal{N}|} )$ is in equilibrium if
\begin{align}\label{eqn:eq_pol}
    J_R(\eta_R^*,\!\{\gamma_i^*(\eta_R^*,\!\gamma_{-i}^*(\eta_R^*))\}_{i=1}^{|\mathcal{N}|} \!) &\!\geq\! J_R (\eta_R,\!\{\gamma_i^*(\eta_R,\!\gamma_{-i}^*(\eta_R))\}_{i=1}^{|\mathcal{N}|} )\nonumber\\
    \gamma_i^*(\eta_R^*,\gamma_{-i}^*(\eta_R)) &\!=\!\mathbf{1}\{ x_{i,ns} \geq L x_S \}, 
\end{align}
where user $i$'s subscription strategy will be based on the term $x_{i,ns}$ affected by the server's strategy and other users' subscription strategies. Thus, at the equilibrium, considering how the users' choose their subscription strategies for a given $\beta$, the server will commit to $\beta^* =\eta_R^*(I_R)$ maximizing its utility, and users adopting $\gamma_i^*$ will not deviate from their action given the server's and other users' subscription strategies. Depending on the network topology, multiple equilibrium solutions may exist. In these scenarios, we assume that the users take the \emph{server-preferred} subscription strategy where the users prefer the equilibrium maximizing the server's utility.\footnote{Another approach could be to consider the worst subscription strategy for the server, and the goal could be to maximize its minimum possible utility.}

\begin{figure}[t]
\centerline{\includegraphics[width=0.73\linewidth]{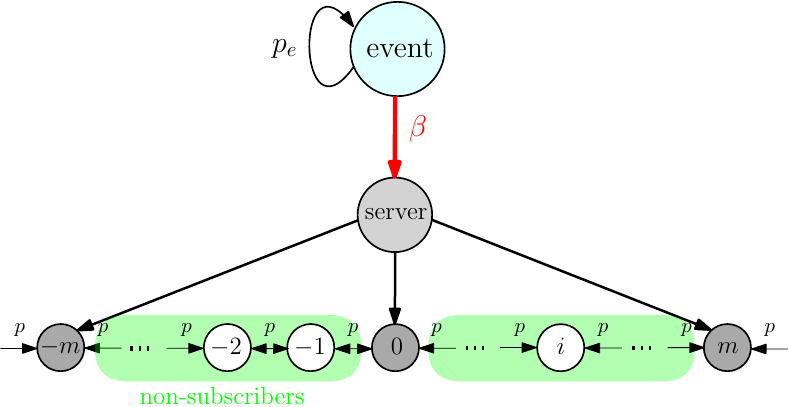}}
\vspace*{-0.2cm}
\caption{An illustration of a typical two-way line network.}
\label{fig:twoway_line_network}
\vspace*{-0.4cm}
\end{figure}

In the next two sections, we delve into subscription policies within two extremes of network connectivity spectrum to provide insightful contrasts on subscription strategies.

\section{Bidirectional Line Network}
We study symmetric AC-stable two-way networks shown in Fig.~\ref{fig:twoway_line_network}, where the number of non-subscribers between any two nearest subscribers is the same. Consider the subscription set $\mathcal{S}=\{mk|k\!\in\!\mathbb{Z}\}$, i.e., subscribers appear with frequency of $m$. Here, $X_i(t+1)$, for any non-subscribing user $i$, depends on the prior age $X_i(t)$ and transitions $U_E(t)$, $U_{i-1,i}(t)$, and $U_{i+1,i}(t)$. If both $U_{i-1,i}(t)\!\!=\!\!1$ and $U_{i+1,i}(t)\!\!=\!\!1$, that is, user $i$ receives update packets from both users $i\!-\!1$ and $i\!+\!1$, it keeps the packet with the lowest instantaneous version age, i.e., $X_i(t+1)=\min\{X_{i-1}(t),X_i(t),X_{i+1}(t)\}$, given $U_E(t)=0$. Likewise, we can evaluate all combinations of the three transitions.

Define $X_{[j,h]}(t)\!\!=\!\!\min\{X_\ell(t)\!\!:\!\!\ell\!\!\in\!\! S_{[j,h]}\}$ to be the instantaneous version age of set of users $S_{[j,h]}\!=\!\{j,\ldots,h\}$, $\forall~0\!\leq\!j \leq$ $h\!\leq\!m$, with $\!x_{[j,h]}\!=\!\!\lim_{t \!\to \!\infty} \!\mathbb{E}[X_{[j,h]}(t)]$. Note that for $j\!=\!0$ or $h\!=\!m$, $x_{[j,h]}=x_S$, since  subscribers have strictly lower instantaneous version age than any non-subscribing user. For other cases, $x_{[j,h]}\! >\! x_S$ and $X_{[j,h]}(t\!+\!1)$ depends on the prior age $X_{[j,h]}(t)$ and transitions $U_{\!E}(t)$, $U_{j-1,j}\!(t)$, $U_{h+1,h}\!(t)$ as,
\begin{align}\label{eqn:xjh_recursiveeqn_allcases}
    x_{[j,h]}=&  x_{[j-1,h]} (1-p_e)(1-p)p+x_{[j,h+1]} (1 \! - \! p_e)p(1 \! - \! p)\nonumber\\
    & + x_{[j-1,h+1]} (1-p_e)p^2+x_{[j,h]} (1-p_e)(1-p)^2\nonumber\\
    &  +(x_{[j-1,h]} \! + \! 1) p_e(1 \! - \! p)p+(x_{[j,h+1]} \! + \! 1) p_ep(1 \! - \! p)\nonumber\\
    &  +(x_{[j-1,h+1]} \! + \! 1) p_ep^2+(x_{[j,h]}+1) p_e(1 \! - \! p)^2,\!\!
\end{align}
which transforms into (with $x_{[i,i]}=x_i$),
\begin{align}\label{eqn:xjh_recursiveeqn}
    x_{[j,h]}\!=& \left(p_e  \! + \! x_{[j-1,h+1]} p^2  \! + \! x_{[j-1,h]} (1 \! - \! p)p \right. \nonumber\\
    &\left.+x_{[j,h+1]} p(1\!-\!p)\right) \!\times\! \left(1\!-\!(1\!-\!p)^2\right)^{-1}
\end{align}

\begin{figure}[t]
\centerline{\includegraphics[width=0.65\linewidth]{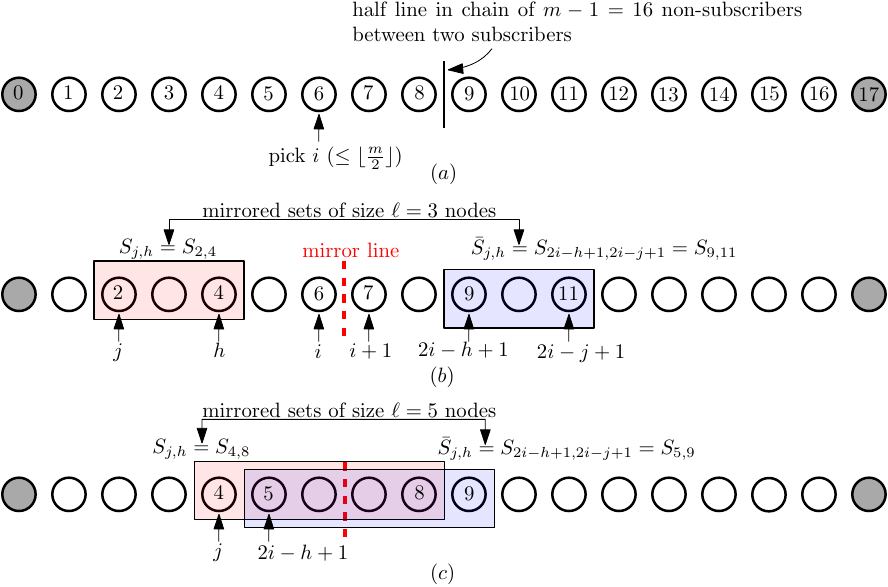}}
\vspace*{-0.4cm}
\caption{$S_{j,h}$ (pink blocks) and $\bar{S}_{j,h}=S_{2i-h-1,2i-j-1}$ (blue blocks) positioned symmetrically about the dotted line between nodes $i$ and $i+1$.}
\label{fig:mirror_2wayline}
\vspace*{-0.5cm}
\end{figure}

\begin{lemma}\label{lemm:highestage_middle_linenetwork}
    The highest age in the non-subscribers $S_{[1,m-1]}$ is at nodes mid-point between subscribers, i.e., $\lfloor \frac{m}{2} \rfloor$ and  $\lceil \frac{m}{2} \rceil$.
\end{lemma}
\begin{Proof}
We show that $x_i\leq x_{i+1}$, $\forall i \leq \lfloor \frac{m}{2} \rfloor$ by induction. Fix an  $i \leq \lfloor \frac{m}{2} \rfloor$ and draw a dotted line between user $i$ and user $i+1$, as shown in Fig.~\ref{fig:mirror_2wayline}. We consider sets $S_{[j,h]}$ with size $\ell=h-j+1$, where at least half of the users fall on the left side of the dotted line (in $S_{[1,i]}$). Define $\Bar{S}_{[j,h]}$ as the mirror image of the set $S_{[j,h]}$ about the dotted line, with $\Bar{S}_{[j,h]}=S_{[2i-h+1,2i-j+1]}$, and the expected age of this set as $\Bar{x}_{[j,h]}=x_{[2i-h+1,2i-j+1]}$. We claim that $x_{[j,h]}\leq \Bar{x}_{[j,h]}$ for all sets of sizes $\ell \in \{2i+1,\ldots,1\}$ with $0\leq j \leq i+1-\frac{\ell}{2}$.

We begin with the biggest sets of $\ell=2i+1$ where the only admissible set is $S_{0,2i}$ with $x_{[0,2i]}=x_S \leq \bar{x}_{[0,2i]}$, satisfying the claim.  
Further, for set size $\ell=2i$, the admissible sets are: (a) $S_{1,2i}=\bar{S}_{1,2i}$, giving $x_{[1,2i]}=\bar{x}_{[1,2i]}$ which satisfies the claim, and (b) $S_{0,2i-1}$ with $x_{[0,2i-1]}=x_S \leq \bar{x}_{[0,2i-1]}$ which satisfies the claim. Next, we show the general inductive step.

We assume that the claim holds for all admissible sets of sizes $\ell$ and $\ell+1$, and prove the claim for all admissible sets of size $\ell-1$. The version age of size $\ell-1$ set depends on version ages of size $\ell$ and $\ell+1$ sets by (\ref{eqn:xjh_recursiveeqn}) as follows,
\begin{align}
    x_{[j,j+\ell-2]}=& \left(p_e  \! + \! x_{[j-1,j+\ell-1]} p^2  \! + \! x_{[j-1,j+\ell-2]} (1 \! - \! p)p \right.\nonumber\\
    &\left.+x_{[j,j+\ell-1]} p(1-p)\right) \times \left(1-(1-p)^2\right)^{-1}  \label{eqn:x_j,j+l-2}\\
    \!\!\Bar{x}_{[j,j+\ell-2]}\!=& x_{[2i-j-\ell+3,2i-j+1]}= \left(p_e  \! + \! \Bar{x}_{[j-1,j+\ell -1]} p^2  \right.\nonumber\\
    &\hspace{-1.75cm}+\! \left.(\Bar{x}_{[j,j+\ell-1]} +\Bar{x}_{[j-1,j+\ell-2]}) p(1\!-\!p)\right) \!\! \times \!\!\left(1\!-\!(1\!-\!p)^2\right)^{\!-1} \!\!\!\!.\!\!\label{eqn:bar_x_j,j+l-2}
\end{align}
The right side of (\ref{eqn:x_j,j+l-2}) and (\ref{eqn:bar_x_j,j+l-2}) deals with size $\ell$ and $\ell \! + \!1$ sets. Hence, by assumptions of the inductive step, we have the terms of (\ref{eqn:x_j,j+l-2}) smaller than terms of (\ref{eqn:bar_x_j,j+l-2}), resulting in $ x_{[j,j+\ell-2]} \leq \Bar{x}_{[j,j+\ell-2]}$. Setting $\ell=1$ and $j=i$ gives $x_i\leq x_{i+1}$.
\end{Proof}

Let $x_i^{(m)}$ and $x_{[j,h]}^{(m)}$ denote the expected age of user $i$ and set $S_{[j,h]}$, respectively, given subscription set $\mathcal{S}=\{mk|k\in \mathbb{Z}\}$.

\begin{lemma}\label{lemm:same_age_diffsets}
     $x_{[j_1,m-h_1]}^{(m)}$ is independent of $m$ for fixed $j_1,h_1$.
\end{lemma}

\begin{figure}[t]
\centerline{\includegraphics[width=0.6\linewidth]{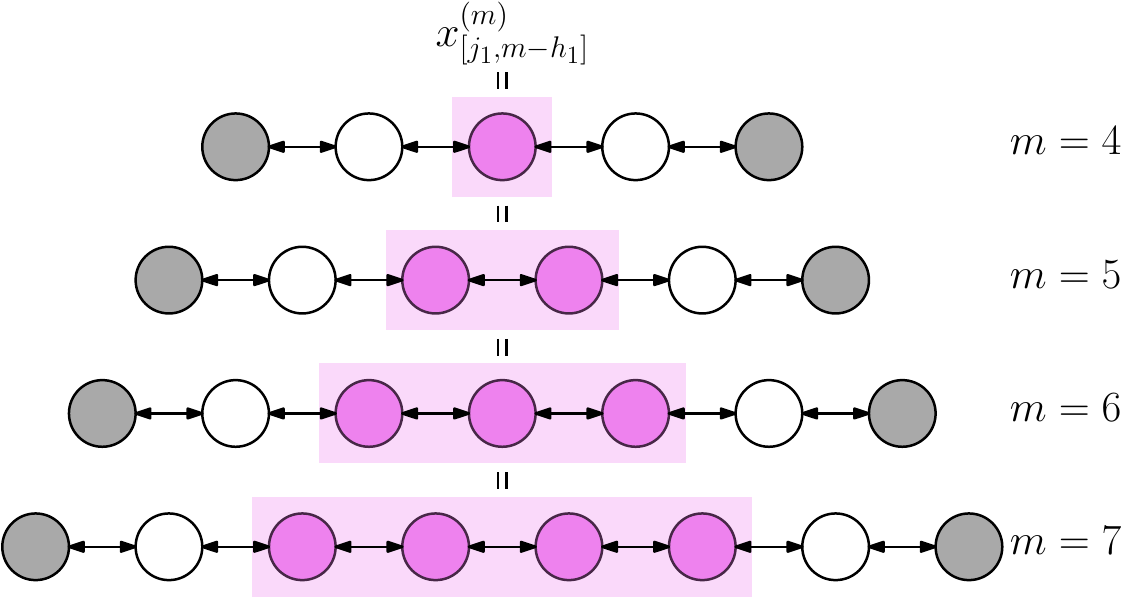}}
\caption{Sets $\{j_1,\ldots,m-h_1\}$ with $j_1=2$ and $h_1=2$, have the same expected age irrespective of the value of $m$.}
\label{fig:same_age_diffsets}
\vspace*{-0.4cm}
\end{figure}

\begin{Proof}
    The set $S_{[j_1,m-h_1]}$ has the same distance from the nearest left and the right subscriber for every $m$; see Fig.~\ref{fig:same_age_diffsets}. Hence, the age processes of these sets are statistically similar, resulting in identical iterative equations from (\ref{eqn:xjh_recursiveeqn}).
\end{Proof}

\begin{lemma}\label{lemm:xm_by_2_increaseswith_m}
We have: (a) $x_{\lfloor \frac{m}{2} \rfloor}^{(m)}$ is an increasing function of $m$. (b) $x_{\lfloor \frac{m}{2} \rfloor}^{(m)}\Big|_{m=1}=x_S$. (c) $\lim_{m\to \infty}x_{\lfloor \frac{m}{2} \rfloor}^{(m)}=\infty$.
\end{lemma}

\begin{Proof}
    The proof involves showing that $x_{\lfloor \frac{m}{2} \rfloor}^{(m)} \leq x_{\lfloor \frac{m+1}{2} \rfloor}^{(m+1)}$. This follows from the fact that the set $S_{[\lfloor \frac{m}{2} \rfloor,\lceil \frac{m}{2} \rceil]}$ with subscription set $\mathcal{S}=\{mk|k\in\mathbb{Z}\}$ and the set $ S_{[\lfloor \frac{m}{2} \rfloor,\lceil \frac{m}{2} \rceil+1]}$ with subscription set $\mathcal{S'}=\{(m+1)k|k\in\mathbb{Z}\}$ have the same distance from the nearest subscribers on both sides, and therefore, by Lemma~\ref{lemm:same_age_diffsets}, these sets have the same expected age. Further $\lfloor \frac{m+1}{2} \rfloor$ belongs to set $S_{[\lfloor \frac{m}{2} \rfloor,\lceil \frac{m}{2} \rceil+1]}$, and the instantaneous age of each node in this set cannot be smaller than the overall age of the set. Hence, $x_{\lfloor \frac{m+1}{2} \rfloor}^{(m+1)} \geq x_{[\lfloor \frac{m}{2} \rfloor,\lceil \frac{m}{2} \rceil+1]}^{(m+1)}=x_{\lfloor \frac{m}{2} \rfloor}^{(m)} $. 

    Next, $m=1$ implies all users are subscribers, hence expected age at all users is $x_S$. Further, even with the best case scenario of $p=1$ when updates immediately get forwarded on all links, the expected age at the middle user from \cite[(13)]{kaswan_subscription_direct} would be $x_0+\lfloor \frac{m}{2} \rfloor p_e \to \infty$ as $m \to \infty $.
\end{Proof}

\begin{lemma}\label{lemm:xi_form_xS_gppe}
    $x_i^{(m)}$ is of the form $x_S+p_eg_1(p,i,m)$, where $g_1(\cdot)$ is a function of gossip probability $p$, node index $i$, and frequency of subscribers $m$ and is independent of $x_S$ and $p_e$.
\end{lemma}

\begin{Proof}
 We claim that $x_{[j,h]}^{(m)}=x_S+p_e g_2(p, j,h,m)$ for all
     $ 0\leq j, h \leq m$, where $g_2(\cdot)$ is a function of $j$, $h$, $p$ and $m$, and is independent of $x_S$ and $p_e$. The claim can be shown for sets $S_{[j,h]}$ of size $m+1$ and $m$, in which case $x_{[j,h]}^{(m)}=x_S$ due to the presence of subscribers in the set, resulting in $g_2(p, j,h,m)=0$. Next, similar to proof of Lemma~\ref{lemm:highestage_middle_linenetwork}, we inductively prove the claim for admissible sets of smaller sizes using (\ref{eqn:xjh_recursiveeqn}).
    
      Finally, note that $g_1(p, i,m)=g_2(p, i,i,m)$ can be obtained by computing $g_2(p, j,h,m)$ from (\ref{eqn:xjh_recursiveeqn}) as follows, 
      \begin{align}\label{eqn:g2_recursion}
      g_2&(p, j,h,m) \!=  \left(1 \!  \! - \!  \! (1\!-\!p)^2\right)^{ \!  \! -1} \! \!  \!  \times \!  \! \big(1 \! +\! g_2(p, j\!-\!1,h\!+\!1,m) p^2  \nonumber\\
      &\hspace{-0.2cm}+ \!g_2(p, j\!-\!1,h,m) (1 \! - \! p)p \!+\!g_2(p, j,h\!+\!1,m) p(1\!-\!p)\big)\!\!\!
      \end{align}
      starting with size $m+1$ and $m$ sets, progressively evaluating for smaller sets upto the singleton set with $j=h=i$.
\end{Proof}

Next, for a given $\beta$, we determine values of $m$ for which the subscription set $\mathcal{S}=\{mk|k\in\mathbb{Z}\}$ guarantees all the users to be AC-stable in the network. We first verify that the subscribers are AC stable. Subscribing user $0$ is AC-stable, if the alternate decision to not subscribe results in an alternate expected age $\Tilde{x}_0$ which violates the AC constraint. Without subscription, user $0$ would depend on updates all the way from subscribers $-m$ and $m$, giving 
\begin{align} \label{eqn:subscriber_cond1}
    \Tilde{x}_0 = x_{m}^{(2m)} \geq Lx_S.
\end{align}

From Lemma~\ref{lemm:xm_by_2_increaseswith_m}, $x_{m}^{(2m)}$ increases with $m$, $\lim_{m\to \infty}x_{m}^{(2m)}=\infty$. Hence, there exists a finite lower bound $m^*=\min\{m:x_{m}^{(2m)} \geq Lx_S,m \in \mathbb{N}\}$. From Lemma~\ref{lemm:xi_form_xS_gppe}, $x_{m}^{(2m)}=x_S+p_e g_1(p,m,2m)$, with $x_S=p_e(\beta^{-1}+1)$. Therefore, $g_1(p,m,2m)$ increases with $m$ and (\ref{eqn:subscriber_cond1}) becomes
\begin{align} \label{eqn:subscriber_cond2}
    p_eg_1(p,m,2m) \geq (L-1)x_S = (L-1)p_e\left(\beta^{-1}+1\right).
\end{align}
Consequently, the lower bound can be defined as
\begin{align} \label{eqn:m_lower_bound}
    \!\! m^* \!=\! \min \left\{m \! : \! g_1(p,m,2m) \!  \geq \! (L-1) \! \left(\!\beta^{-1} \! + \! 1 \! \right), m \! \in \! \mathbb{N} \right\} \!
\end{align}
such that, for $m \geq m^*$, the subscribing users will be AC-stable. Next, we verify that the non-subscribing users are AC-stable, which would provide an upper-bound $m^{**}$ on $m$. From Lemma~\ref{lemm:highestage_middle_linenetwork}, it suffices to satisfy AC constraint for user $\lfloor m/2\rfloor$,
\begin{align}\label{eqn:nonsubscriber_cond1}
    x_{\lfloor \frac{m}{2} \rfloor}^{(m)} < Lx_S.
\end{align}
Define $m^{**}=\max\{m:x_{\lfloor \frac{m}{2} \rfloor}^{(m)} < Lx_S\}$. Then, Lemma~\ref{lemm:xi_form_xS_gppe} gives
\begin{align}\label{eqn:m_upper_bound}
    \!\!\! m^{**} \! = \! \max\left\{m \! : \! g(p, \! \lfloor  \! \frac{m}{2} \! \rfloor \!,m) \!  < \! (L \! - \! 1) \! \left(\beta^{-1} \! + \! 1\right) \! ,m \! \in  \! \mathbb{N}\right\}\!\!.\!\!
\end{align}
Observe from (\ref{eqn:m_lower_bound}) and (\ref{eqn:m_upper_bound}) that both $m^{*}$ and $m^{**}$ decrease with increase in $\beta$, allowing for more frequent subscribers.

\begin{lemma} 
For a given $\beta$, the symmetric bidirectional network is AC-stable $\forall ~ m\in\{m^*,\ldots,m^{**}\}$, with $m^*=\left\lfloor m^{**}/2\right\rfloor+1$.
\end{lemma}

\begin{Proof}
    The last part follows from the fact that if we choose $m=2m^*$, then the central non-subscriber $\lfloor \frac{m}{2} \rfloor$ will not be AC-stable due to expected age $x_{\lfloor \frac{m}{2} \rfloor}^{(m)}=x_{m^*}^{(2m^*)}  \geq Lx_S$ from intial definition of $m^*$. Therefore, $m\leq m^{**}<2m^*$. 
    
    Conversely, for $m=2m^*-2$, all non-subscribers are AC-stable, implying $m^{**}\geq 2m^{*}-2$. This is because, the  central non-subscriber $\lfloor \frac{m}{2} \rfloor$ is AC-stable and will not subscribe, as in the latter case, subscribers will appear at a frequency of $m^*-1$ and will not be AC-stable by definition of $m^*$. 
\end{Proof}

In this work, among these multiple AC-stable networks, we choose the symmetric AC-stable network with the maximum number of subscribers, i.e.,  by $S^* = \{m^*k|k\in\mathbb{Z} \}$.\footnote{As we mentioned earlier, another approach might be to consider the AC-stable network that gives the lowest utility. In that case, we would choose the subscription set as $S^{**} = \{m^{**}k|k\in\mathbb{Z} \}$ and follow the same steps.} Thus, we have $F_{S,2{\text -}line}(\beta) = \frac{1}{m^*}$ where $m^*$, defined in (\ref{eqn:m_lower_bound}), depends on $\beta$. The minimum $\beta=\beta^*(m)$ required to maintain a subscription frequency of $m$, using equality in (\ref{eqn:m_lower_bound}), is $\beta^*(m)=\left( \frac{g_1(p,m,2m)}{(L-1)} -1 \right)^{-1}\!\!\!\!,$ where $g_1(p,m,2m)=g_2(p,m,m,2m)$ is obtained by solving (\ref{eqn:g2_recursion}) recursively as explained in  Lemma~\ref{lemm:xi_form_xS_gppe}. Next, we use the server's minimum $\beta^*(m)$ to find $\beta^*_{2{\text -}line}$ that maximizes the server's utility, that is, $\beta^*_{2{\text -}line} = \argmax_{\beta^*(m),m\in\mathbb{N}}  F_{S,2{\text -}line}(\beta^*(m))- c(\beta^*(m)).$

\section{Fully-Connected Networks}
\begin{figure}[t]
\centerline{\includegraphics[width=0.45\linewidth]{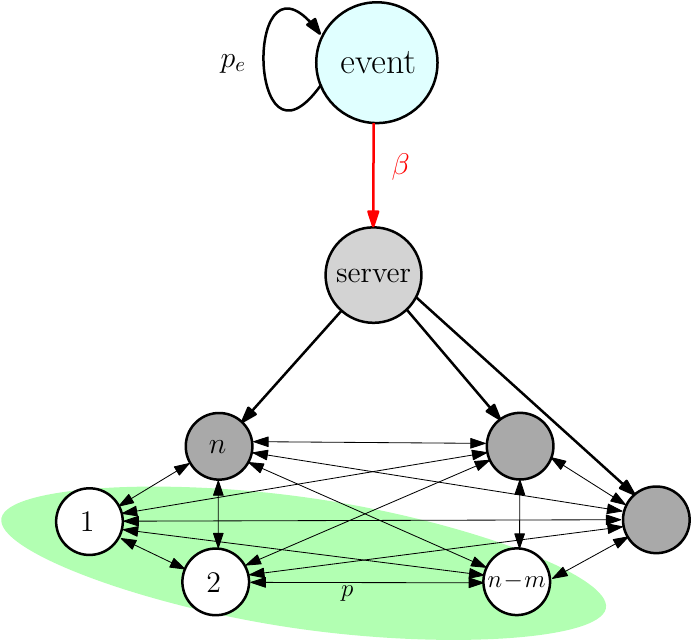}}
\vspace*{-0.2cm}
\caption{A typical  fully-connected network with $m$ subscribers.}
\label{fig:fully_connected_subscriber}
\vspace*{-0.4cm}
\end{figure}

In a fully-connected network, each of the total $n$ users transmits an update to every other user in each timeslot with probability $p$; see Fig.~\ref{fig:fully_connected_subscriber}. There are $m$ subscribers and $n-m$ non-subscribers, with the subscription set as $\mathcal{S}=S_{[n-m+1,n]}$.

Let $x_{NS}^{(m)}$ denote expected age at a typical non-subscribing user in a fully-connected network with $m$ subscribers. Due to network symmetry, every set of $k$ non-subscribing users will have the same expected age, which we represent by $x_{[1,k]}^{(m)}$. To analytically characterize $x_{[1,k]}^{(m)}$, we consider how combinations of various transitions affect $X_{[1,k]}(t)$ evolution. The set $S_{[1,k]}$ has $k$ non-subscribers, and it can externally receive update packets from the $m$ subscribers and the remaining $n-m-k$ non-subscribers. For the case of no event update in timeslot $t$, the probability of not receiving an update from any subscriber is $(1-p)^{km}$, thus with probability $1-(1-p)^{km}$, $X_{[1,k]}(t+1)=X_S(t)$. In the absence of any updates from subscribers, the probability of receiving updates from exactly $i$ out of the remaining $n-m-k$ non-subscribers is $\binom{n-m-k}{i}(1-(1-p)^k)^i (1-p)^{k(n-m-k-i)}$. For the case of updates arriving from the $i$ non-subscribers in $S_{[k+1,k+i]}$, we will have $X_{[1,k]}(t+1)=X_{[1,k+i]}(t)$. Accounting for all possible cases, we get 
\begin{align}\label{eqn:age_recursion_fullyconnec}
        \!x_{[1,k]}^{(m)}\!= &\frac{1}{1\!-\!(1\!-\!p)^{k(n\!-\!k)}} \times \bigg[p_e + \sum_{i=1}^{n\!-\!m\!-\!k}\binom{n\!-\!m\!-\!k}{i} x_{[1,k+i]}^{(m)} \nonumber\\
        & \hspace{-0.9cm}\times \! \!(1-(1-p)^k)^i  (1\!-\!p)^{k(n\!-\!k\!-\!i)} +  (1\!-\!(1\!-\!p)^{km})x_S  \bigg] .
\end{align}
We can use (\ref{eqn:age_recursion_fullyconnec}) to iteratively compute $x_{[1,k]}^{(m)}$ in the order $k=n-m,\ldots,1$, to finally yield $x_{[1,1]}^{(m)}=x_{1}^{(m)}=x_{NS}^{(m)}$.

\begin{lemma}\label{lemm:x1k_form_g_fullconn}
    (a) $x_{[1,k]}^{(m)}$ is in the form of $x_S+p_eg(p,m,k)$. \\ (b) $x_{[1,k]}^{(m)}$ and $g(p,m,k)$ are decreasing functions of $m$.
\end{lemma}

\begin{Proof}
    Similar to Lemma~\ref{lemm:xi_form_xS_gppe}, the first part can be proven by induction, starting from $k=n-m$ and iteratively proving for $k=n-m-1,\ldots,1$ using (\ref{eqn:age_recursion_fullyconnec}). Similar to (\ref{eqn:g2_recursion}), $g(p,m,k)$ can be computed iteratively for $k=n-m,\ldots,1$ using the following equation obtained from (\ref{eqn:age_recursion_fullyconnec}),
     \begin{align}
        &g(p,\!m,\!k) \!=\! \bigg[1+\sum_{i=1}^{n-m-k} \!\!\! \binom{n-m-k}{i} g(p,m,k+i)  \nonumber\\
        & \!\!  \times  \!\! (1 \! - \! (1\!-\!p)^k)^i (1\!-\!p)^{k(n-k-i)}  \! \bigg] \!\!  \times\!\!\left(1 \! - \! (1 \!\! - \! p)^{k(n-k)}\right)^{ \!\! -1} 
    \end{align}
    To prove the second part, we claim $x_{[1,k]}^{(m+1)}\leq x_{[1,k]}^{(m)}$, $\forall k\leq n-m$, which we prove inductively in the order $k=n-m,\ldots,1$. Assuming the claim holds for non-subscribing sets of size greater than $k$, we can prove the claim for set size $k$ by comparing terms of $x_{[1,k]}^{(m)}$ and $x_{[1,k]}^{(m+1)}$ (obtained from (\ref{eqn:age_recursion_fullyconnec}) by substituting $m+1$ in place of $m$), where $\binom{n-m-k}{i} \geq \binom{n-m-k-1}{i}$, $(1-(1-p)^{km})\geq (1-(1-p)^{k(m+1)})$, and $x_{[1,k+i]}^{(m)} \geq x_{[1,k+i]}^{(m+1)}$, and $x_{[1,k]}^{(m)}$ in has an additional summation term when $i=n-m-k$ in (\ref{eqn:age_recursion_fullyconnec}). The proof for $g(p,m,k)$ then follows from the first part.
\end{Proof}
\begin{lemma}
For a given $\beta$, there exists a unique $m=m^*_{FC}$, for which the fully-connected network is AC-stable with $m^*_{FC}$ subscribers and $n-m^*_{FC}$ non-subscribers, such that,
\begin{align}\label{eqn:mfully_m*_equality}
       m^*_{FC}=\max\{m:x_S+p_eg(p,m-1,1) \geq Lx_S\}. 
    \end{align} 
\end{lemma}

\begin{Proof}
For non-subscribers to be AC-stable, we need $x_{1}^{(m)}<Lx_S$. From Lemma~\ref{lemm:x1k_form_g_fullconn} part (b),  $x_{1}^{(m)}$ decreases with $m$, with $ x_{1}^{(m)}|_{m=0}=\infty >Lx_S$ and $ x_{1}^{(m)}|_{m=n}=x_S< Lx_S$. Hence, there exists a unique $m=m^{*}_{FC}$, such that, 
\begin{align}\label{eqn:fullyconnec_bothcondn}
          x_{1}^{(m^{*}_{FC})}< Lx_S \leq x_{1}^{(m^{*}_{FC}-1)} .
\end{align}
For $m \!  < \!  m^{*}_{FC}$, the non-subscribers will not be AC-stable from the right inequality of (\ref{eqn:fullyconnec_bothcondn}) and Lemma~\ref{lemm:x1k_form_g_fullconn}(b). For $m \!> \! m^{*}_{FC}$, any subscriber will not be AC-stable, and will unsubscribe and still satisfy AC constraint with the rest $m-1$ subscribers from the left inequality of (\ref{eqn:fullyconnec_bothcondn}) and Lemma~\ref{lemm:x1k_form_g_fullconn}(b). Thus, $m^*_{FC} \! = \! \max\{m \! : \! x_{1}^{(m-1)}  \!  \! = \! x_S \! + \! p_eg(p,m \! - \! 1,1)  \! \geq  \! Lx_S\}$, from combining right inequality of (\ref{eqn:fullyconnec_bothcondn}) and Lemma~\ref{lemm:x1k_form_g_fullconn}(a).
\end{Proof}

Hence, $F_{S,FC}(\beta)  =\frac{m^*_{FC}}{n}$, where $m^*_{FC}$ depends on $\beta$ from (\ref{eqn:mfully_m*_equality}). The minimum $\beta=\beta^*(m)$ required to maintain $m$ subscribers, from equality in (\ref{eqn:mfully_m*_equality}), is  $\beta^*(m)=\left( \frac{g(p,m-1,1)}{(L-1)} -1 \right)^{-1}$, which can be used to determine the server's optimum $\beta^*_{FC}$ via, $\beta^*_{FC} = \argmax_{\beta^*(m),m =1,2,\cdots}  F_{S,FC}(\beta^*(m))- c(\beta^*(m)).$

\section{Simulation Results}
\begin{figure}[t]
\centerline{\includegraphics[width=0.95\linewidth]{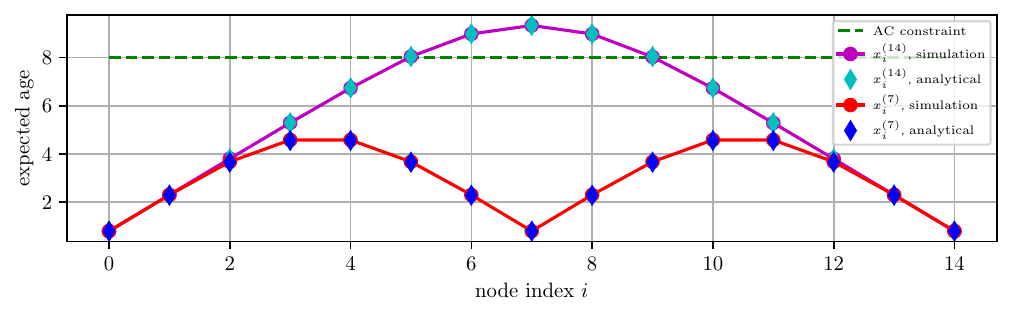}}
\vspace*{-0.3cm}
\caption{The expected age of nodes in a two way line network with $m\!=\!\{7\!,\!14\}\!$.}
\label{fig:graph_xi_vs_i_shortestcell_twowayLine_numerical_VS_analytical}
\vspace*{-0.4cm}
\end{figure}

We simulate the bidirectional line network in Fig.~\ref{fig:graph_xi_vs_i_shortestcell_twowayLine_numerical_VS_analytical} with parameters $L  \! \! = \! \!  10$, $p \! \!  =  \! \! 0.2$, $\beta \!  \! =  \! \! 0.6$, $p_e \! \!  =  \! \! 0.3$ for a total time of $t \!  \! = \!  \! 10^4$, which we use as a proxy for $t \to \infty$, and take the average over $2\times10^5$ iterations to approximate $\lim_{t  \!  \! \to \!  \!  \infty} \mathbb{E}[X_i(t)]$ by the law of large numbers. The real-time simulation points (red) coincide with the analytically derived values (blue) from (\ref{eqn:xjh_recursiveeqn_allcases}), verifying the theoretical analysis. Observe how all non-subscribers are AC-stable with $m^*=7$, i.e., $x_i^{(7)}$ is below the AC constraint threshold (green) for all $i$, and the subscribers are AC-stable, since the alternate decision to unsubscribe increases their age to $x_7^{(14)}$, which violates the AC constraint. 

\begin{figure}[t]
\centerline{\includegraphics[width=0.54\linewidth]{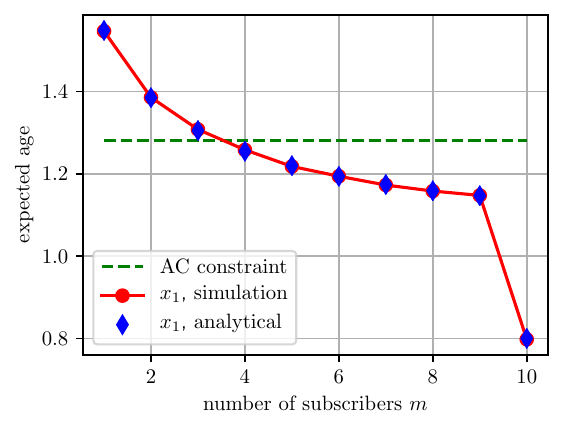}}
\vspace*{-0.4cm}
\caption{The expected age of a node in a fully-connected network of $n=10$.} 
\label{fig:graph_x1_vs_m_fullyconnec_numerical_VS_analytical}
\vspace*{-0.2cm}
\end{figure}

\begin{figure}[t]
\centerline{\includegraphics[width=0.5\linewidth]{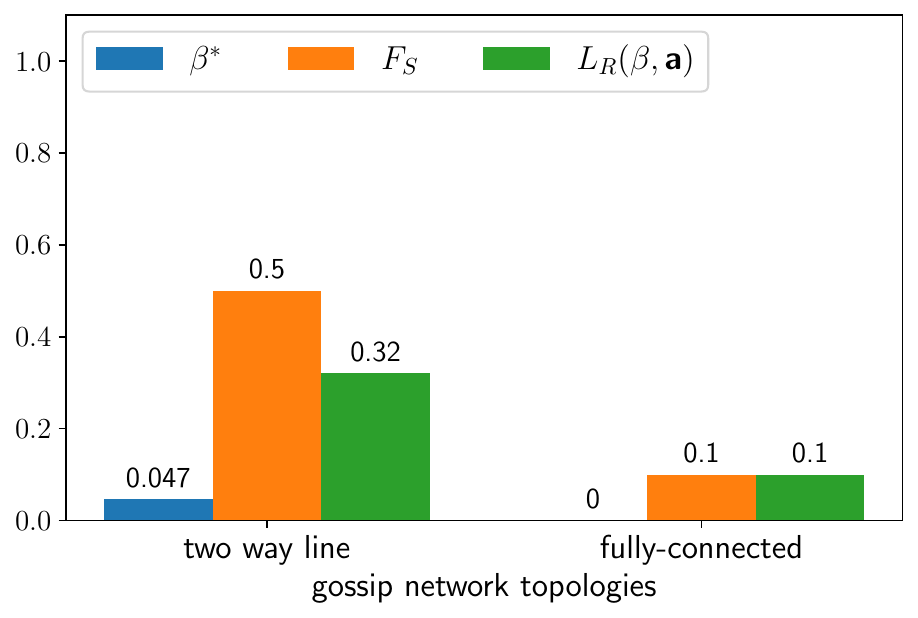}}
\vspace*{-0.4cm}
\caption{The comparison of the server's optimal $\beta^*$, subscription rate $F_S$, and the server's optimal cost function for two way-line network}
\label{fig:graph_graph_beta_FS_LS_undirected}
\vspace*{-0.5cm}
\end{figure}

Next we focus on expected age of user $1$ in fully connected network of $n \! = \! 10$ users, with $L\!=\!1.6$, $p\!=\!0.2$, $\beta \! =\!0.6$, $p_e\!=\!0.3$. The decrease of $\!x_1^{(10)}\!\!$ with increase in $m$ is in accordance with Lemma~\ref{lemm:x1k_form_g_fullconn}(b), with $m^*_{FC}=4$ making all users AC-stable, by (\ref{eqn:mfully_m*_equality}). The sudden jump in Fig.~\ref{fig:graph_x1_vs_m_fullyconnec_numerical_VS_analytical} at $\!m\!=\!n\!=\!10\!$ comes from user $1$ itself turning to a subscriber. Fig.~\ref{fig:graph_graph_beta_FS_LS_undirected} shows $\beta^*$, the corresponding $F_S$, and server utility function $L_R(\beta, \bm{a})$ in both networks with $L\!=\!1.6$, $p\!=\!0.2$, $\beta\!=\!0.6$, $p_e\!=\!0.3$ and server utility function $F_S(\beta) \! - 80 \!  \times \! \beta^2$ which penalizes high $\beta$, but is neutral to small $\beta$ due to the $80 \times\beta^2$ term. The deep connectivity of the fully-connected network disincentivizes users from subscribing, resulting in just one subscriber and $\beta^*_{2{\text -}line}\to 0=\beta^{*}(1)$ that suffices to sustain one subscriber. Contrarily, the bidirectional line has every alternate user has subscriber, with higher $\beta^*_{FC}$, $F_S$ and utility. 

\section{Conclusion}
We investigated a Stackelberg game between a server and users, where the server wishes to maximize its profit by increasing subscribers and reducing event sampling costs, and the users base their subscription decisions on the server strategy and their timeliness requirements. In the bidirectional line network, multiple Stackelberg equilibrium solutions exist, with higher plausible periodicity of subscribers under frequent server sampling. The fully-connected network gives a single equilibrium solution and fewer subscribers owing to the deep network connectivity disincentivizing users from subscribing.

\bibliographystyle{unsrt}
\bibliography{ref_priyanka}

\end{document}